%Paper: hep-ph/9506287
%From: jelwood@theory.caltech.edu (John Elwood)
%Date: Fri, 9 Jun 95 18:38:25 PDT

\input phyzzx.tex
\input epsf.tex

% Titlepage macros
\hoffset=0.2truein
\voffset=0.1truein
\hsize=6truein
\def\TITLEPAGE{\frontpagetrue}
\def\CALT#1{\hbox to\hsize{\tenpoint \baselineskip=12pt
        \hfil\vtop{
        \hbox{\strut CALT-68-#1}}}}

\def\AUTHOR#1{\vskip .2in \centerline{#1}}

\def\ABSTRACT#1{\vskip .2in \vfil \centerline{\twelvepoint
\bf Abstract}
        #1 \vfil}
\def\ENDTITLEPAGE{\vfil\eject\pageno=1}

\def\K{K_L\rightarrow \pi^+ \pi^- e^+ e^-}

\def\k{\hbox{/\kern-.5600 em {$k$}}}

\TITLEPAGE
%\CALT{1993}
\hskip 4.3in{\vbox{\hbox{CALT-68-1993}
\hbox{CMU-HEP95-04}
\hbox{DOE-ER/40682-95}}}
\bigskip            %if title takes 2 lines use \break
\titlestyle {\bf Final State Interactions and CP Violation in $K_L \rightarrow
\pi^+ \pi^- e^+ e^-$}
\AUTHOR{{\bf John K. Elwood and Mark B. Wise}\foot{Work
 supported in part  by U.S. Department of Energy Grant
 No. DE-FG03-92-ER40701.}}
\medskip
\centerline{\it Lauritsen Laboratory, California Institute of Technology,
Pasadena, CA 91125}
\medskip
\AUTHOR{{\bf Martin J. Savage and James W. Walden}\foot{Work
 supported in part by U.S. Department of Energy Grant No.
 DE-FG02-91-ER40682.}}
\medskip
\centerline{\it Department of Physics, Carnegie Mellon University, Pittsburgh,
PA 15213}
\ABSTRACT
{Using chiral perturbation theory we calculate the imaginary parts of the $K_L
\rightarrow \pi^+ \pi^- e^+ e^-$ form factors that arise from $\pi\pi
\rightarrow \pi^+\pi^-$ and $\pi\pi \rightarrow \pi^+\pi^- \gamma^*$
rescattering.  We discuss their influence on CP violating variables in
$K_L \rightarrow \pi^+ \pi^- e^+ e^-$.}

\ENDTITLEPAGE

\eject

The E832 fixed target experiment at Fermilab, whose primary goal is to look for
a nonzero value of $\epsilon^\prime/\epsilon$, will reconstruct on the order of
1000 events in the rare decay mode $K_L \rightarrow \pi^+ \pi^- e^+ e^-$[1].
At present, approximately 10 such events have been observed by the E731 fixed
target experiment [2], the precursor to E832.  Long-distance physics dominates
this decay mode, with the leading contribution coming from $K_L \rightarrow
\pi^+ \pi^- \gamma^* \rightarrow \pi^+ \pi^- e^+ e^-$, where a single virtual
photon creates the $e^+e^-$ pair.  This one photon contribution to the decay
amplitude has the form
$$
M^{(1\gamma)} = {s_1 G_F\alpha\over 4\pi fq^2} [i G \epsilon^{\mu\lambda
\rho\sigma} p_{+\lambda}  p_{-\rho} q_\sigma + F_+ p_+^\mu + F_- p_-^\mu] \bar
u (k_-) \gamma_\mu v (k_+) ,\eqno (1)
$$
where $G_F$ is Fermi's constant, $\alpha$ is the electromagnetic fine structure
constant, $s_1 \simeq 0.22$ is the sine of the Cabibbo angle and $f \simeq$ 132
MeV is the pion decay constant.  The $\pi^+$ and $\pi^-$ four momenta are
denoted by $p_+$ and $p_-$ while the $e^+$ and $e^-$ four momenta are denoted
by $k_+$ and $k_-$.  The sum of electron and positron four-momenta is $q = k_+
+ k_-$.  The Lorentz scalar form factors $G, F_\pm$ depend on scalar products
of the four momenta $q, p_+$ and $p_-$.  Theoretical predictions for
$G$,$F_\pm$ were first made in Ref. [3].

Chiral perturbation theory allows a systematic expansion of an observable in
powers of $p^2$, where $p$ is a typical momentum involved in the process of
interest.  Such an expansion was performed for the form factors $F_\pm$ and $G$
defined above in the analysis of Ref. [4]:

$$
F_\pm = F_\pm^{(1)} + F_\pm^{(2)} +...$$
$$ \quad G = G^{(1)} + G^{(2)} + ... \eqno (2)$$
The superscripts denote the order of chiral perturbation theory at which each
term arises (i.e., $F_{\pm}^{(m)}$, $G^{(m)}$ give a contribution of order
$p^{2m-1}$ to the square brackets of eq. (1))

The $K_L$ state has both CP even and CP odd components
$$
|K_L> \simeq |K_2> + \epsilon |K_1> \eqno (3)
$$
where $|K_2>$ is the CP odd state $|K_2> = (|K^0> + |\bar K^0>)/\sqrt{2}$ and
$|K_1>$ is the CP even state $|K_1> = (|K^0> - |\bar K^0>)/\sqrt{2}$.  The
parameter $\epsilon \simeq 0.0023 e^{i44^{0}}$ (in a phase convention where the
$K^0 \rightarrow \pi\pi (I = 0)$ amplitude is real) characterizes  CP
nonconservation in $K^0 - \bar K^0$ mixing.  We neglect other (i.e., direct)
sources of CP nonconservation in the one photon part of the $\K$ decay
amplitude.  Contributions to the form factors $F_\pm$ from the $|K_2>$ and
$|K_1>$ parts of the $K_L$ state have different symmetry properties.  Under
interchange of the pion four momenta, $p_+ \rightarrow p_-$ and $p_-
\rightarrow p_+$, the CP conserving parts of the form factors arising from the
$|K_2>$ component transform as
$$
F_+ \rightarrow F_-,{~~\rm and~~} F_- \rightarrow F_+, \eqno (4)
$$
while the CP violating parts of the form factors arising from the $|K_1>$
component transform as
$$
F_+ \rightarrow - F_-, {~~ \rm and~~} F_- \rightarrow - F_+. \eqno (5)
$$

At leading order in chiral perturbation theory (i.e., order $p$ in the square
brackets of eq. (1))
$$
G^{(1)} = 0 \eqno (6a)
$$
$$
F_+^{(1)} = - {32g_8 f^2 (m_K^2 - m_\pi^2) \pi^2 \epsilon\over q^2 + 2q \cdot
p_+} \eqno (6b)
$$
$$
F_-^{(1)} = {32 g_8 f^2 (m_K^2 - m_\pi^2) \pi^2 \epsilon\over q^2 + 2q \cdot
p_-} \eqno (6c)$$
$G^{(1)}$ is zero (it enters in the square brackets of eq. (1) multiplied by
three momentum factors, and is therefore at most an order $p^3$ effect) and
contributions to $F_\pm$ not proportional to $\epsilon$ don't occur until
higher order in chiral perturbation theory.  In eqs.~(6), $g_8$ is the
coefficient of the leading two-derivative part of the chiral Lagrangian for
$\Delta S = 1$ weak nonleptonic kaon decay [5].  It is real and the measured
$K^0 \rightarrow \pi\pi (I = 0)$ decay amplitude gives $|g_8| \simeq 5.1$.

Since the CP violating contribution to the $\K$ decay amplitude occurs at a
lower order of chiral perturbation theory than the CP conserving contribution,
the effects of indirect CP nonconservation are enhanced in this decay.  It is
convenient for the discussion of CP violation in $\K$ to use the four-body
phase space variables used by Pais and Trieman for semileptonic $K_{\ell 4}$
decay [6].  They are: $q^2 = (k_+ + k_-)^2$; $s = (p_+ + p_-)^2$; $\theta_\pi$,
the angle between $\pi^+$ three-momentum and the $K_L$ three-momentum in the
$\pi^+\pi^-$ rest frame; $\theta_e$, the angle between the $e^-$ three-momentum
and the  $K_L$ three-momentum in the $e^+e^-$ rest frame and; $\phi$, the angle
between the normals to the planes defined (in the $K_L$ rest frame) by the
$\pi^+\pi^-$ pair and the $e^+e^-$ pair.  Using these kinematic variables the
CP violating observable
$$
B_{CP} = <sign (\sin \phi \cos \phi)> \eqno (7)
$$
gets a large contribution from indirect CP nonconservation.  Neglecting other
sources of CP violation, one has, after integrating over $\cos\theta_e$ and
$\phi$
$$
B_{CP} = {G_F^2 s_1^2 \alpha^2\over 3 \cdot 2^7 (2\pi)^8 f^2 m_K^3
\Gamma_{K_{L}}} \int d \cos \theta_\pi \, ds \, dq^2 \, \sin^2 \theta_\pi
\beta^3 X^2 \left({s\over q^2}\right) Im [G(F_+^* - F_-^*)]. \eqno (8)
$$
where
$$\beta = [1 - 4m^2_{\pi}/s]^{1/2} \eqno (9a)
$$
$$X = \left [ \left (m^2_K - s -q^2 \over 2 \right )^2 - s q^2 \right ]^{1/2}
\eqno (9b)
$$
If the variables $s$ and $q^2$ are not integrated over the entire phase space
then the same is to be done to the $\K$ width, $\Gamma_{K_{L}}$, in the
denominator of eq. (8).

The form factor $G$ first arises at second order in chiral perturbation theory.
 Because tree diagrams involving vertices from the Wess--Zumino term don't
contribute~[7], it is dominated by local order $p^4$ terms in the chiral
Lagrangian [8] which give a real contribution to $G^{(2)}$.  The measured $K_L
\rightarrow \pi^+ \pi^- \gamma$ decay rate [9] implies that
$$
|G^{(2)}| \simeq 40. \eqno (10)
$$
In obtaining this result from the data, we have neglected the experimental
momentum dependence of $G$.  Higher order terms in the chiral expansion endow
$G$ with momentum dependence.  At leading order in chiral perturbation theory
$$
Im [G(F_+^* - F_-^*)] \rightarrow Im [G^{(2)} (F_+^{(1)*} - F_-^{(1)*})], \eqno
(11)
$$
in eq. (8) and the imaginary part comes solely from the phase of $\epsilon$
appearing in $F_\pm$.  In Ref.~[3] the form factors $F_\pm$ and $G$ were
estimated by extrapolating from the measured $K_L \rightarrow \pi^+ \pi^-
\gamma$ amplitude.  They noted that $B_{CP}$ was large and furthermore showed
that final state $\pi\pi$ interactions give an important enhancement  of
$B_{CP}$.  In this letter we calculate the absorptive parts of $G$ and $(F_+ -
F_-)$ using chiral perturbation theory and consider their influence on
$B_{CP}$.  Our approach includes both $\pi\pi \rightarrow \pi^+ \pi^-$ and
$\pi\pi \rightarrow \pi^+ \pi^- \gamma^*$ rescattering.  Previous estimates of
the effect of final state interactions used the measured pion phase shifts and
neglected $\pi\pi \rightarrow \pi^+\pi^- \gamma^*$.

Dividing the third order contribution to $G$ into its dispersive and absorptive
pieces, $G^{(3)} = Disp G^{(3)}  + iAbs G^{(3)}$, we find that the Feynman
graph shown in Fig.~1 gives
$$ Abs G^{(3)} = {G^{(2)}\over 48\pi} \left( {s\over f^2}\right) \left( 1-
{4m^2 _{\pi}\over s} \right)^{3/2}. \eqno (12)$$
Unfortunately, the dispersive part of $G^{(3)}$ is not calculable as it
receives a contribution not only from the loop graph in Fig. 1, but also from
loop graphs involving the Wess--Zumino term and from new order $p^6$ local
operators in the chiral Lagrangian for weak radiative kaon decay.

The absorptive parts of $F_\pm$ first arise at second order in chiral
perturbation theory from the Feynman diagrams in Figs. 2  which give
$$
AbsF_+^{(2)} = - g_8 (m^2_K - m^2_{\pi}) \pi \epsilon \bigg \{{(4m_K^2 -
2m_\pi^2)\over q^2 + 2q \cdot p_+} \sqrt{1-{4m_\pi^2\over m_K^2}}$$
$$
- 4 \bigg [\int_0^{\xi_{-}} y_+ dy - \int_0^{\xi_{+}} y_- dx\bigg ] $$
$$
- {8q \cdot (p_+ - p_-)\over s} \bigg[ \int_0^{\xi_{+}} {xy_-\over (y_+ - y_-)}
dx + \int_0^{\xi_{-}} {xy_+\over (y_+ - y_-)} dx \bigg] \bigg \}. \eqno (13)$$
$Abs F_-^{(2)}$ is obtained from eq. (13) by interchanging $p_+$ with $p_-$
using the symmetry property in eq. (5).  The limits of integration in eq. (13)
are given by
$$
\xi_\pm = {1 \pm \sqrt{1-4m_\pi^2/m_K^2}\over 2}, \eqno (14)$$
and the variables $y_\pm$ are defined by
$$
y_\pm = {(1-x)s + x (m_K^2 - q^2) \pm \sqrt{((1 - x) s + x (m_K^2 - q^2))^2 -
4s (m_\pi^2 - q^2 x (1-x))}\over 2s}. \eqno (15)$$
We include the influence of final state interactions on $B_{CP}$ by setting
$$
Im [G (F_+ - F_-)^*] \rightarrow Im [G^{(2)} (F_+^{(1)} - F_-^{(1)})^*]$$
$$
+ Re [Abs G^{(3)} (F_+^{(1)} - F_-^{(1)})^*] - Re [G^{(2)} (Abs F_+^{(2)} - Abs
F_-^{(2)})^*] , \eqno (16)$$
in eq. (8).  The first of the three terms on the right hand side of eq. (16)
was calculated in Ref. [4] and the last two represent the effects of final
state interactions.

We find that final state interactions increase $B_{CP}$ by about 45\% over what
we presented in Ref. [4].  The first term in eq. (13), and consequently the
third term in eq. (16), is the dominant contribution from final state
interactions and it enhances $B_{CP}$ by the factor
$${(4m_K^2 - 2m_\pi^2)\over 32\pi f^2} \sqrt{1-{4m_\pi^2\over m_K^2}} \simeq
0.45 \eqno (17)$$
over the leading order result obtained in Ref. [4]. The trend that final state
interactions increase $B_{CP}$ is in agreement with Ref. [3]. The rate
$\Gamma_{K_L}$ in the denominator of equation (8) depends on the collection of
counterterms defined as $w_L$ in Ref. [4]. Setting $w_L$ to zero, we find that
$|B_{CP}| \simeq $ 14\% with the cut $q^2 > (10 MeV)^2$ imposed and $|B_{CP}|
\simeq $ 4\% with the cut $q^2 > (80 MeV)^2$ imposed. With $w_L=2$, the
asymmetry is even larger.  We find in this case that $|B_{CP}| \simeq $ 18\%
for each of the cuts listed above.  Table 1 gives the predicted values for the
magnitude of $B_{CP}$ times the branching ratio for $\K$ (in units of
$10^{-8}$) for various cuts on the minimum lepton pair invariant mass squared,
$q^2_{min}$.

\medskip
\input tables
\begintable
{\bf Lower cut $q^2_{min}$} | $\mid B_{CP}(\%)\mid \cdot Br(10^{-8})$~\cr
$(~10 MeV)^2$ | 208\cr
$(~20 MeV)^2$ | 122\cr
$(~30 MeV)^2$ | ~76\cr
$(~40 MeV)^2$ | ~50\cr
$(~60 MeV)^2$ | ~22\cr
$(~80 MeV)^2$ | 9.7\cr
$(100 MeV)^2$ | 3.9\cr
$(120 MeV)^2$ | 1.4\cr
$(180 MeV)^2$ | 0.013\endtable
\centerline { {\bf Table 1}: The CP violating observable $|B_{CP}| \cdot
Br(10^{-8})$}
\centerline { for a range of values of $q^2_{min}$.}

We have calculated the leading absorptive parts of the form factors $G$ and
$F_\pm$ using chiral perturbation theory and included, using eq. (16), their
influence on $B_{CP}$.  However, this is not a completely systematic approach
because $Im[DispG^{(3)} (F_+^{(1)} - F_+^{(1)})^*]$ and $Im[G^{(2)}
(DispF^{(2)}_+ - DispF^{(2)}_-)^*]$ in eq. (16) were neglected, despite being
the same order in the momentum expansion as the terms that were retained.
Nonetheless, including only the absorptive parts may be a good approximation as
they are enhanced by a factor of $\pi$.

Finally we note that the absorptive parts of the form factors calculated here
are also important for direct CP nonconservation in $\K$.  For example, the
variable
$$
D_{CP} = <sign(\cos \theta_e)>, \eqno (18)$$
is a CP violating observable that arises from interference of the one photon
amplitude in eq. (1) with the short distance contribution to the $\K$ decay
amplitude,
$$
M^{(SD)} = {G_F s_1 \alpha\over f} (\xi p_-^\mu + \xi^* p_+^\mu ) \bar u
(k_-)\gamma_\mu \gamma_5 v (k_+) . \eqno (19)$$
In the kaon rest frame, the electron-positron energy difference is proportional
to $\cos \theta_e$;  $D_{CP}$ is therefore a measure of this $e^+e^-$ energy
asymmetry.

The W-box and Z-penguin Feynman diagrams are responsible for producing the
short distance amplitude, $M^{(SD)}$.  The quantity $\xi$ depends on the charm
and top quark masses and on Cabibbo--Kobayashi--Maskawa matrix elements.  It
has been calculated in the next to leading logarithmic approximation [10].
After integrating over $\phi$ and $\cos \theta_e$ we find that
$$
D_{CP} = {s_1^2 G_F^2 \alpha^2\over 2^7 (2\pi)^6 m_K^3 f^2 \Gamma_{K_{L}}} \int
d\cos\theta_\pi \,  ds \, dq^2 \, \beta^3 X^2 \sin^2\theta_\pi s Im G Im \xi .
\eqno  (20)$$
At leading order in chiral perturbation theory $ImG =  Abs G^{(3)}$.
Unfortunately, we find that $D_{CP}$ is too small to be measured in the next
generation of kaon decay experiments and for this reason do not present
numerical results for it here.

In this work, we have estimated the final state interactions at lowest order in
the chiral expansion for strong interactions.  Higher order contributions which
we have not computed may modify our results, particularly in the I=J=1 channel
where the $\rho$ plays an important role [11].

In summary, we have determined the leading effect of $\pi\pi\rightarrow\pi\pi$
and $\pi\pi\rightarrow\pi\pi\gamma^*$ final state interactions on the CP
violating asymmetry $B_{CP} = \langle sign(\sin\phi \cos\phi) \rangle$.  We
find that these interactions enhance $B_{CP}$ by about 45\% over the estimates
given in~[4].  We have also shown that the CP violating $e^+e^-$ energy
asymmetry $D_{CP}$  arises from the interference of the short-distance
amplitude with the absorptive part of the form factor G, but found that
$D_{CP}$ is unlikely to be observed in the near future.

\bigskip
\centerline{\bf Acknowledgements}
\medskip

MJS would like to thank the Institute for Nuclear Theory at the University of
Washington for its kind hospitality during the course of this work. He would
also like to thank Barry Holstein for useful discussions.

\vfill
\eject

\leftline{\bf References}
\vskip 0.5in

\item{1.}  Y.W. Wah, private communication.
\item{2.}  Y.W. Wah, in {\it Proceedings of the XXVI International Conference
on High Energy Physics}, Dallas, Texas, AIP Conf. Proc. No. 272, edited by
James R. Sanford (AIP, New York) , 1992.
\item {3.}  L.M. Sehgal and M. Wanninger, {\it Phys. Rev.} {\bf D46} (1992)
1035; P. Heiliger and L.M. Sehgal, {\it Phys. Rev.} {\bf D48} (1993) 4146.
\item{4.}  J.K. Elwood, M.B. Wise and M.J. Savage, CALT-68-1980 (1995)
unpublished.
\item {5.} G. Ecker, A. Pich, and E. de Rafael, {\it Nucl. Phys.} {\bf B291}
(1987) 692.
\item {6.} A. Pais and S. Trieman, {\it Phys. Rev.} {\bf 168} (1968) 1858.
\item {7.} J. Kambor, J. Missimer, and D. Wyler {\it Nucl. Phys.} {\bf B346}
(1990) 17; J. Bijnens, G. Ecker, and A. Pich, {\it Phys. Lett.} {\bf B286}
(1992) 341;  G. D'Ambrosio and G. Isidori, {\it Z. Phys.} {\bf C65} (1995) 649.
\item {8.} G. Ecker, J. Kambor, and D. Wyler, {\it Nucl. Phys.} {\bf B394}
(1993) 101.
\item {9.} E. J. Ramberg et al., {\it Phys. Rev. Lett.} {\bf 70} (1993) 2525.
\item {10.} G. Buchalla and A. J. Buras, {\it Nucl. Phys.} {\bf B398} (1993)
1285;
ibid {\bf B400} (1993) 225.
\item {11.}  J. Donoghue, C. Ramirez, and G. Valencia, {\it Phys. Rev.}\ {\bf
D39} (1989) 1947.
\vfill
\eject

\centerline {\bf Figure Captions}
\item {\rm Fig. 1:} Feynman diagram contributing to $AbsG^{(3)}$ at leading
order.  In this figure and those that follow, a solid circle denotes a vertex
arising from the leading order strong and electromagnetic chiral Lagrangian.
The other vertex in this figure arises from an $O(p^4)$ counterterm in the
chiral Lagrangian.
\smallskip
\item {\rm Fig. 2:} Feynman diagrams contributing to $AbsF^{(2)}_{\pm}$ at
leading order.  A solid square denotes a vertex arising from the $\Delta S = 1$
part of the leading order gauged weak chiral Lagrangian.  A solid triangle
vertex arises from the piece of the leading order strong chiral Lagrangian
proportional to the quark masses.
\vfill
\eject
%\nopagenumbers
\vbox{\vskip -3truecm
\centerline{\epsffile{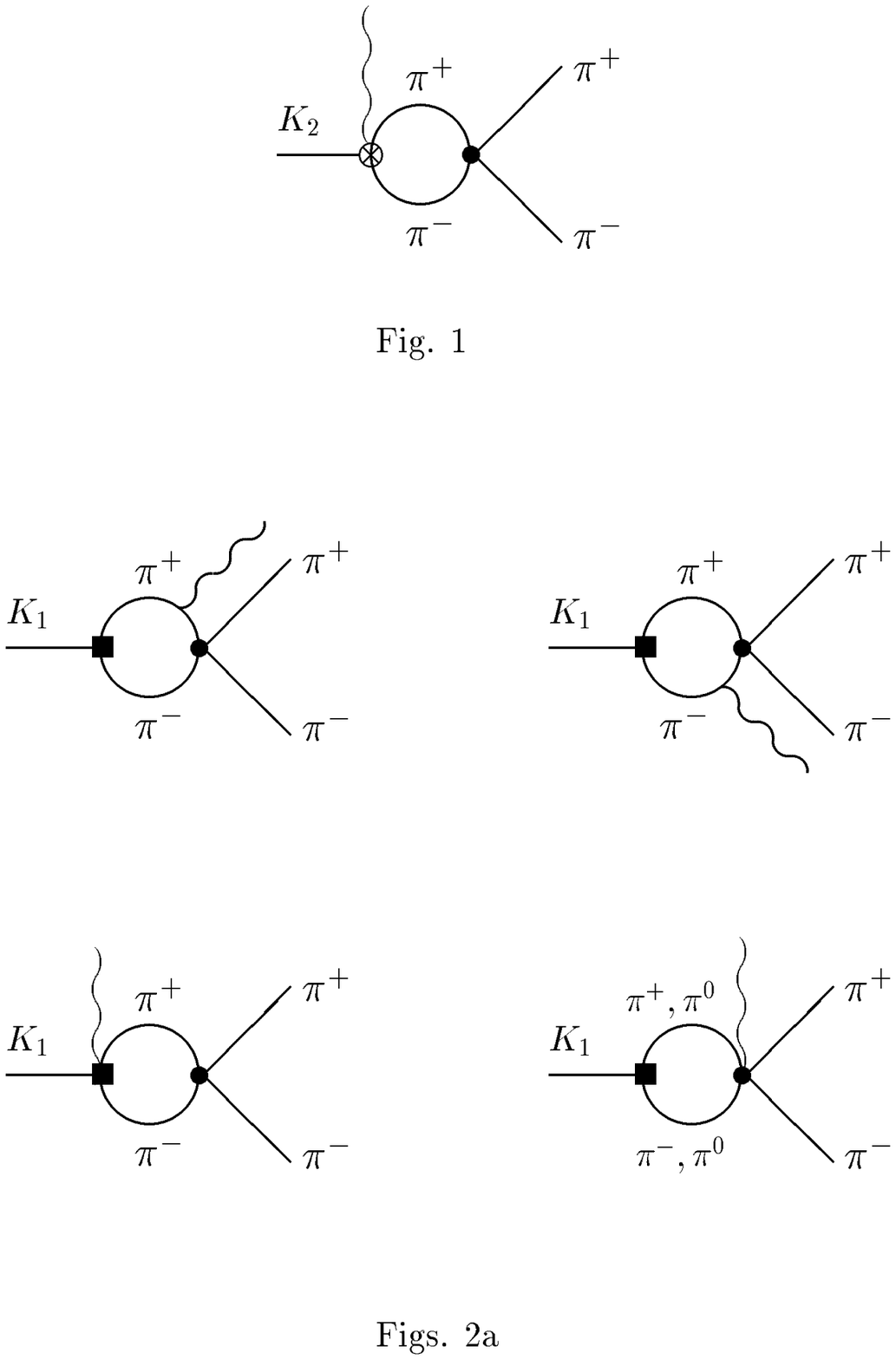}} }
\vfill
\eject
%\nopagenumbers
\vbox{\vskip -3truecm
\centerline{\epsffile{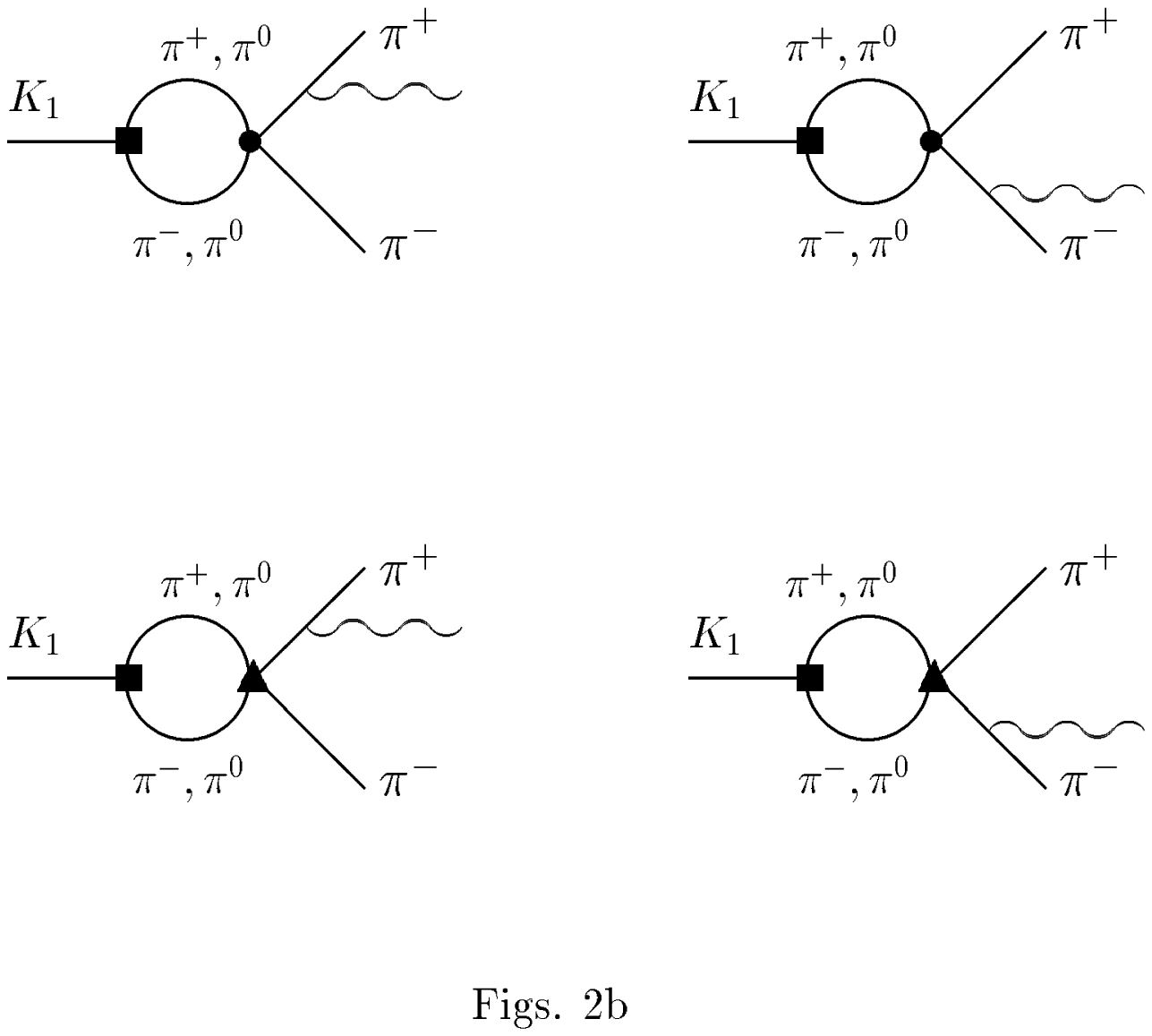}} }

\bye